\documentclass{article}
\usepackage{spconf,amsmath,amssymb,graphicx}
\usepackage[T1]{fontenc}
\usepackage[utf8]{inputenc}
\usepackage{cite,url}
\usepackage{booktabs}
\usepackage{tikz}
\usetikzlibrary{arrows.meta, positioning, shadows, fit, backgrounds, calc}
\usepackage{fontawesome5}
\usepackage{color}
\usepackage{makecell}
\usepackage{multirow}
\usepackage[bookmarks=false,hidelinks]{hyperref}
\usepackage{xcolor}
\usepackage{xurl}
\definecolor{my_blue}{RGB}{50, 110, 210}
\definecolor{my_green}{RGB}{70, 160, 90}
\definecolor{my_red}{RGB}{200, 70, 70}
\definecolor{my_gray}{RGB}{100, 100, 100}
\definecolor{frozen_blue}{RGB}{100, 180, 255}
\definecolor{my_orange}{RGB}{230, 130, 40}

\newcommand{\tf}[2]{%
    {\color{black!75}#1/#2}%
}
\newsavebox{\audiowavebox}
\sbox{\audiowavebox}{%
	\begin{tikzpicture}[x=0.15cm, y=0.15cm, baseline=-0.5ex]
		\draw[thick, my_blue!80!black] (0,0) -- (1,2) -- (2,-2) -- (3,3) -- (4,-2.5) -- (5,2) -- (6,-1.5) -- (7,1) -- (8,0);
	\end{tikzpicture}%
}

\newsavebox{\watermarkedwavebox}
\sbox{\watermarkedwavebox}{%
	\begin{tikzpicture}[x=0.15cm, y=0.15cm, baseline=-0.5ex]
		\draw[thick, my_blue!80!black] (0,0) -- (1,2) -- (2,-2);
		\draw[very thick, my_red] (2,-2) -- (3,3) -- (4,-2.5) -- (5,2) -- (6,-1.5);
		\draw[thick, my_blue!80!black] (6,-1.5) -- (7,1) -- (8,0);
	\end{tikzpicture}%
}

\newsavebox{\distortedwavebox}
\sbox{\distortedwavebox}{%
	\begin{tikzpicture}[x=0.15cm, y=0.15cm, baseline=-0.5ex]
		\draw[thick, my_red!60!black] (0,0) -- (0.8,2.2) -- (1.2,1.6) -- (2.2,-2.2) -- (2.8,-1.6) -- (3.2,3.2) -- (4,-2.8) -- (4.8,-2.0) -- (5.2,2.2) -- (6,-1.8) -- (7,1.2) -- (8,0);
	\end{tikzpicture}%
}

\tikzset{
	box/.style={
		rectangle, draw=my_gray, thick, fill=white, align=center,
		rounded corners=3pt, drop shadow={opacity=0.2}
	},
    codecanchor/.style={
    rectangle, draw=none, fill=none,
    text=my_green!60!black, font=\bfseries,
    align=center,
    minimum width=2.4cm, minimum height=1.9cm
    },
    attacksub/.style={
        rectangle, draw=my_red!70, thick, fill=my_red!10,
        text=my_red!50!black, font=\small\bfseries,
        minimum width=2.6cm, minimum height=0.55cm,
        rounded corners=2pt, align=center
    },
    zoneclean/.style={
        rectangle, draw=my_blue!50, semithick,
        dash pattern=on 3pt off 2pt,
        fill=my_blue!4, rounded corners=8pt, inner sep=11pt
    },
    zonerecover/.style={
        rectangle, draw=my_red!50, semithick,
        dash pattern=on 3pt off 2pt,
        fill=my_red!4, rounded corners=8pt, inner sep=11pt
    },
	data/.style={box, draw=my_blue, fill=my_blue!5, text=my_blue!60!black, font=\bfseries, minimum width=2.8cm, minimum height=1.9cm},
	model/.style={box, draw=my_green, fill=my_green!5, text=my_green!60!black, font=\bfseries, minimum width=2.4cm, minimum height=1.4cm},
	attack/.style={box, draw=my_red, fill=my_red!5, text=my_red!60!black, font=\bfseries, dashed, minimum width=2.4cm, minimum height=1.4cm},
	arrow/.style={->, >={Stealth[scale=1.0]}, thick, draw=my_gray!80, shorten >=2pt, shorten <=2pt},
	lossarrow/.style={<->, >={Stealth[scale=1.0]}, thick, draw=my_orange, dashed, rounded corners=5pt, shorten >=2pt, shorten <=2pt},
	regarrow/.style={->, >={Stealth[scale=1.0]}, thick, draw=my_orange, dashed, rounded corners=5pt, shorten >=2pt, shorten <=2pt},
	arrowtext/.style={font=\scriptsize\itshape, text=my_gray}
}

\renewcommand{\arraystretch}{0.9}

\title{Latent Audio Watermarking for Robustness to Neural Codec Resynthesis}

\name{Lovro Brulec$^{1,2}$\sthanks{Equal contribution.}, Sahil Karawade$^{1,\ast}$, Leonard Kinzinger$^{1,2}$}

\address{
$^1$Munich Music Labs\quad$^2$Technical University of Munich\\
{\small\texttt{lovro.brulec@tum.de, sahil.karawade@munichmusiclabs.com, leonard.kinzinger@tum.de}}
}

\begin{document}
\ninept

\maketitle

\begin{abstract}
    Existing waveform-domain audio watermarks are robust to many conventional distortions but can degrade substantially under neural codec resynthesis. We investigate whether continuous neural codec latents provide a more suitable embedding space using a restricted formulation built around frozen pretrained EnCodec. To test this, a feedforward embedder maps a multi-bit payload to an additive latent perturbation decoded through the unchanged codec decoder.
    
    Compared with AudioSeal and WavMark, our latent watermark formulation degrades more gradually under repeated and low-bitrate EnCodec resynthesis, transfers to unseen DAC, and retains high detection under most waveform distortions. Substantial EnCodec robustness emerges even without codec-resynthesis supervision, indicating that this behavior is inherent to our latent formulation and is further strengthened by codec-aware training. Learned perturbations are also preserved more strongly through codec cycling than equal-norm random controls, with preservation depending more on channel-specific allocation than temporal structure. End-to-end perceptual quality remains close to that of the frozen EnCodec reconstruction, indicating that much of the observed degradation originates from the codec carrier itself. Overall, these results show that continuous neural codec latents provide a promising embedding space for watermarks that remain robust to neural codec resynthesis.
\end{abstract}
\begin{keywords}
    Audio watermarking, neural audio codecs, latent representations, robustness
\end{keywords}
\section{Introduction}\label{sec:introduction}

    As synthetic speech and music become indistinguishable from human-produced recordings, reliable mechanisms for provenance, attribution, and misuse detection are becoming a pressing technical need. Audio watermarking has therefore attracted growing interest as a means of embedding authenticity or ownership information directly into the signal~\cite{Madiega2023Watermarking}. A practical watermark, however, must remain imperceptible, reliably decodable, and robust to downstream transformations~\cite{Salah2024}.

    Established methods such as AudioSeal~\cite{AudioSeal}, WavMark~\cite{WavMark}, SilentCipher~\cite{SilentCipher}, and Timbru~\cite{timbru} satisfy these requirements for most conventional distortions but cannot be reliably recovered after neural codec resynthesis~\cite{Rawbench,AudioMarkBench}. Neural codec resynthesis poses a particular challenge because watermarking methods aim to embed imperceptible information, while neural codecs are designed to discard information that is unnecessary for perceptually faithful reconstruction~\cite{Rawbench}. Since neural codecs form the backbone of current generative audio systems, this transformation is moreover among the most likely to be encountered in the very setting that motivates watermarking.
    
    We therefore investigate embedding the watermark closer to the internal representations that neural codecs preserve. Several recent studies explore latent-domain watermarking, for example using optimization-based latent perturbations~\cite{timbru,LatentMark}, discrete intermediate representations~\cite{DiscreteWM}, steganographic shifts in frozen autoencoders~\cite{RoSteALS}, learned continuous latent embedding within codec-like autoencoders~\cite{Hu2026CodecInternal}, and watermarking the training data of latent audio generative models~\cite{LatentWatermarking}. In contrast, we investigate whether a deliberately simple, feedforward perturbation of a frozen neural codec latent can provide robustness to codec resynthesis, and how such robustness compares with established waveform-domain methods in terms of conventional attack resistance and perceptual transparency.

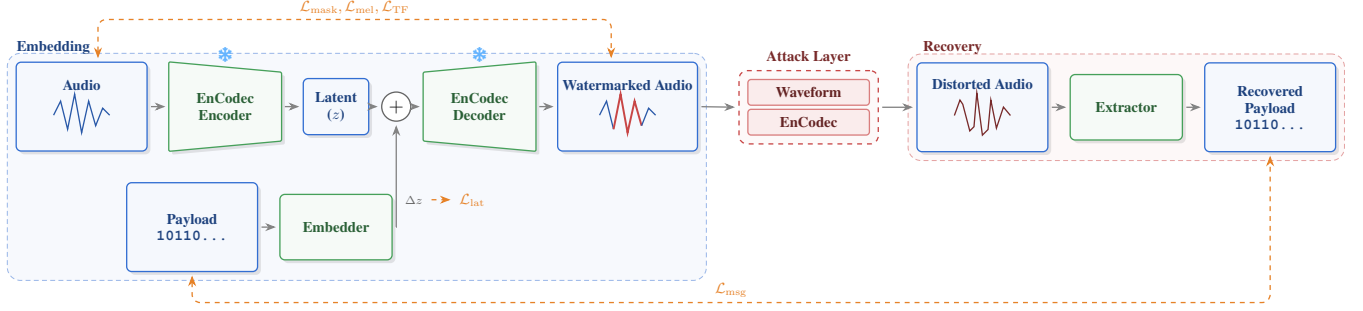
\begin{figure*}[t]
    \centering
    \resizebox{\textwidth}{!}{
        \begin{tikzpicture}[node distance=0.4cm and 0.45cm]

            \node (audio) [data]
                {Audio\\[1pt] \usebox{\audiowavebox}};

            \node (encode) [codecanchor, right=of audio]
                {EnCodec\\Encoder};
            \node[text=frozen_blue, font=\small]
                at ($(encode.north)+(0,0.16)$) {\faSnowflake};

            \node (latent) [data, right=of encode,
                            minimum width=1.4cm, minimum height=1.2cm]
                {Latent\\($z$)};

            \node (plus) [circle, draw=my_gray, thick, inner sep=2pt,
                          font=\large, fill=white, right=0.25cm of latent]
                {$+$};

            \node (decode) [codecanchor, right=0.25cm of plus]
                {EnCodec\\Decoder};
            \node[text=frozen_blue, font=\small]
                at ($(decode.north)+(0,0.16)$) {\faSnowflake};

            \node (watermarked) [data, right=of decode]
                {Watermarked Audio\\[1pt] \usebox{\watermarkedwavebox}};

            \node (attack_wave) [attacksub, right=1.05cm of watermarked, yshift=0.305cm]
                {Waveform};
            \node (attack_neural) [attacksub, below=0.06cm of attack_wave]
                {EnCodec};

            \begin{scope}[on background layer]
                \node[
                    fit=(attack_wave)(attack_neural),
                    draw=my_red, dashed, thick, rounded corners=5pt,
                    fill=my_red!3, inner sep=5pt
                ] (attack_box) {};
            \end{scope}

            \node[
                above=0.01cm of attack_box,
                text=my_red!60!black,
                font=\bfseries\small
            ] {Attack Layer};

            \node (distorted) [data, right=0.8cm of attack_box]
                {Distorted Audio\\[1pt] \usebox{\distortedwavebox}};

            \node (extractor) [model, right=of distorted]
                {Extractor};

            \node (logits) [data, right=of extractor]
                {Recovered\\Payload\\[-1pt] \texttt{10110...}};

            \node (embedder) [model, below=1.25cm of latent]
                {Embedder};

            \node (message) [data, left=of embedder]
                {Payload\\[-1pt] \texttt{10110...}};

            \draw [arrow] (audio) -- (encode);
            \draw [arrow] (encode) -- (latent);
            \draw [arrow] (latent) -- (plus);
            \draw [arrow] (plus) -- (decode);
            \draw [arrow] (decode) -- (watermarked);

            \draw [arrow] (watermarked.east) -- (attack_box.west);
            \draw [arrow] (attack_box.east) -- (distorted.west);

            \draw [arrow] (distorted) -- (extractor);
            \draw [arrow] (extractor) -- (logits);

            \draw [arrow] (message) -- (embedder);

            \draw [arrow] (embedder) -|
                node[pos=0.63, arrowtext, right, xshift=2pt]
                (deltaz) {$\Delta z$}
                (plus);

            \begin{scope}[on background layer]

                \node[
                    fit=(audio)(watermarked)(embedder)(message),
                    zoneclean,
                    inner sep=5pt
                ] (zone_embed) {};

                \node[
                    fit=(distorted)(logits),
                    zonerecover,
                    inner sep=5pt
                ] (zone_recover) {};

                \path[
                    draw=my_green, thick, fill=my_green!5,
                    drop shadow={opacity=0.2}, rounded corners=1pt
                ]
                    (encode.north west)
                    -- (encode.south west)
                    -- ([yshift=0.22cm]encode.south east)
                    -- ([yshift=-0.22cm]encode.north east)
                    -- cycle;

                \path[
                    draw=my_green, thick, fill=my_green!5,
                    drop shadow={opacity=0.2}, rounded corners=1pt
                ]
                    ([yshift=-0.22cm]decode.north west)
                    -- ([yshift=0.22cm]decode.south west)
                    -- (decode.south east)
                    -- (decode.north east)
                    -- cycle;

            \end{scope}

            \node[
                anchor=west,
                font=\small\bfseries,
                text=my_blue!60!black
            ]
                at ([xshift=0.1cm,yshift=0.12cm]zone_embed.north west)
                {Embedding};

            \node[
                anchor=west,
                font=\small\bfseries,
                text=my_red!60!black
            ]
                at ([xshift=0.22cm,yshift=0.12cm]zone_recover.north west)
                {Recovery};

            \draw [lossarrow]
                ([xshift=0.35cm]audio.north)
                -- ++(0,0.9) coordinate(P1)
                -- node[above, text=my_orange, font=\bfseries]
                    {$
                    \mathcal{L}_{\mathrm{mask}},
                    \mathcal{L}_{\mathrm{mel}},
                    \mathcal{L}_{\mathrm{TF}}
                    $}
                ([xshift=-0.35cm]watermarked.north |- P1)
                -- ([xshift=-0.35cm]watermarked.north);

            \draw [lossarrow]
                (message.south)
                -- ++(0,-0.65) coordinate(M1)
                -- node[above, text=my_orange, font=\bfseries]
                    {$\mathcal{L}_{\mathrm{msg}}$}
                (logits.south |- M1)
                -- (logits.south);

            \draw [regarrow]
                (deltaz.east)
                -- ++(0.55,0)
                node[right, text=my_orange, font=\bfseries]
                    {$\mathcal{L}_{\mathrm{lat}}$};

        \end{tikzpicture}
    }

    \caption{
    Latent watermarking training pipeline.
    A payload-dependent perturbation $\Delta z$ is added to the frozen EnCodec latent and decoded to audio.
    Training optimizes payload recovery after attacks under perceptual and latent regularization.
    Snowflakes mark frozen components.
    }
    \label{fig:simplified_pipeline}
\end{figure*}

    We study this question using a minimal post-hoc formulation built around a frozen pretrained EnCodec~\cite{EnCodec} model, illustrated in Fig.~\ref{fig:simplified_pipeline}. A multi-bit payload is mapped to an additive perturbation in the codec's continuous latent representation and decoded through the unchanged codec decoder. The perturbation depends only on the payload, so the same message produces the same latent perturbation regardless of the input audio. This deliberately prevents the embedder from adapting to individual signals and allows us to test whether a shared payload-dependent direction in latent space can survive waveform decoding, re-encoding, and quantization. We refer to this formulation as DeltaMark. It is fully feedforward and requires neither codec retraining nor per-audio optimization.

        Our contributions are:
    \begin{itemize}
        \setlength{\itemsep}{0pt}
        \setlength{\parskip}{0pt}
        \setlength{\parsep}{0pt}
        \setlength{\topsep}{2pt}
    
        \item[(i)] a content-independent latent watermark formulation for a frozen neural codec that requires neither codec retraining nor per-signal optimization;
        \item[(ii)] a comparison with AudioSeal and WavMark under repeated and low-bitrate EnCodec resynthesis, the unseen codecs DAC~\cite{DAC} and TiCodec~\cite{TiCodec}, conventional distortions, and perceptual evaluation, demonstrating stronger robustness under EnCodec resynthesis.
        \item[(iii)] an ablation demonstrating that a substantial part of the EnCodec robustness is obtained without codec-aware training.
        \item[(iv)] a perturbation analysis showing that preservation through EnCodec depends more strongly on channel organization than temporal structure.
    \end{itemize}

    Code is available at \url{https://github.com/Munich-Music-Labs/latent-audio-watermarking}.

\section{Method}\label{sec:method}

    \subsection{Latent watermarking method}
        Let $x \in \mathbb{R}^{T}$ denote an input waveform and $s \in \{0,1\}^{K}$ a binary payload. A frozen EnCodec encoder $E(\cdot)$ produces a continuous latent representation $z=E(x)$. The embedder $F(\cdot)$ predicts a payload-dependent perturbation $\Delta z=\alpha F(s)$, which is added before decoding:
        \[
            x_w = D\!\left(z+\Delta z \right),
        \]
        where $D(\cdot)$ is the frozen codec decoder and $\alpha$ is a learned scalar controlling perturbation magnitude. During training, the watermarked waveform is passed through a stochastic attack operator $A(\cdot)$ and then the extractor $H(\cdot)$ to predict $\hat{s}=H(A(x_w))$.

        We use the official pretrained mono EnCodec model at 24~kHz and bypass residual vector quantization (RVQ) during watermark insertion. For 3-second clips, the encoder produces a $128\times225$ latent representation.

        The embedder is implemented as a three-layer MLP with hidden width 256 and GELU~\cite{Gelu} activations. The 16-bit payload is projected to a $128\times64$ coarse latent tensor and linearly interpolated to the EnCodec temporal length of 225.
    
        The extractor is a 1D residual convolutional network with four stages of widths $[32,64,128,256]$, followed by global average pooling and a linear head that predicts the $K$ payload bits.

    \subsection{Training objective}
        The embedder $F$ and extractor $H$ are trained jointly with an objective that balances payload recoverability under distortions with perceptual transparency:
        \[        \mathcal{L}_{\mathrm{total}}
                =
                \lambda_{\mathrm{msg}} \mathcal{L}_{\mathrm{msg}}
                +
                \lambda_{\mathrm{lat}} \mathcal{L}_{\mathrm{lat}}
                +
                \lambda_{\mathrm{mask}} \mathcal{L}_{\mathrm{mask}}
                +
                \lambda_{\mathrm{mel}} \mathcal{L}_{\mathrm{mel}}
                +
                \lambda_{\mathrm{TF}} \mathcal{L}_{\mathrm{TF}}.
        \]

        Robustness is encouraged through the message loss \mbox{$\mathcal{L}_{\mathrm{msg}} =
        \operatorname{BCE}(\hat{\mathbf{s}},\mathbf{s})$},
        where $\hat{\mathbf{s}}$ is predicted from the attacked waveform $x_a = A(x_w)$. The remaining terms constrain the perceptual impact of watermark insertion. We penalize the energy of the latent perturbation using $\mathcal{L}_{\mathrm{lat}}=\|\Delta z\|_2^2$, together with a log-Mel spectral loss and the time--frequency loudness objective introduced by AudioSeal~\cite{AudioSeal}.
        
        We additionally use a masking-inspired spectral loss~\cite{Swanson1998,Moritz2024NMR} that penalizes residual magnitude more strongly in low-magnitude time--frequency regions. Let $r=x_w-x$ denote the watermark reconstruction residual, with $X(f,t)=|\operatorname{STFT}(x)|$ and \mbox{$R(f,t)=|\operatorname{STFT}(r)|$}, where $\operatorname{STFT}$ denotes the short-time Fourier transform. We define
        \[
            \mathcal{L}_{\mathrm{mask}}
            =
            \mathbb{E}_{f,t}
            \left[
                \frac{R(f,t)}
                {\max(X(f,t),\epsilon)^{\gamma}+\epsilon}
            \right].
        \]
        We average this loss across Hann-window STFTs with FFT/window sizes $[256,512,1024]$ and hops $[64,128,256]$, using $\gamma=0.5$ and $\epsilon=10^{-5}$.

\section{Experimental setup}\label{sec:setup}
    \subsection{Training setup}\label{sec:train_protocol}
        Training uses 16,000 clips from the LibriSpeech train-clean-100 corpus~\cite{LibriSpeech}, with 2,000 additional clips for validation. All audio is converted to mono, resampled to 24~kHz, peak-normalized to 0.95, and segmented into 3-second clips.
    
        The embedder and extractor are trained jointly with frozen EnCodec using online random 16-bit payloads. We use AdamW~\cite{Loshchilov2019AdamW} with batch size 64, learning rate $10^{-3}$, weight decay $10^{-2}$, and a ReduceLROnPlateau scheduler for a maximum of 500 epochs, with early stopping on attacked validation loss after 30 epochs without improvement.

        Following staged optimization used in prior watermarking systems~\cite{WavMark,RoSteALS}, training proceeds through four stages summarized in Table~\ref{tab:stages}. Bootstrap uses a 1,000-clip subset, after which training continues on the full dataset. Stage transitions occur at 80\% and 98\% clean validation bit accuracy for Bootstrap and Generalization, respectively, and at 99.2\% worst-attack validation bit accuracy for Robustness. We reinitialize the optimizer at the Robustness and Refinement stages to avoid carrying adaptive optimizer state across substantial changes in the training objective.

        \begin{table}[h]
            \centering
            \ninept
            \begin{tabular}{@{}lllccccc@{}}
                \toprule
                Stage  & Attacks & $\lambda_{\mathrm{msg}}$ & $\lambda_{\mathrm{mask}}$ & $\lambda_{\mathrm{lat}}$ & $\lambda_{\mathrm{mel}}$ & $\lambda_{\mathrm{TF}}$ \\ \midrule
                Bootstrap            & --  & 8.0 & 0.067 & 0.01 & -- & -- \\
                Generalization       & --  & 1.0 & 0.4   & 0.01 & -- & -- \\
                Robustness           & yes & 1.0 & 0.667 & 0.06 & -- & -- \\
                Refinement           & yes & 1.0 & 0.667 & 0.06 & 0.2 & 0.2 \\ 
                \bottomrule
            \end{tabular}
            \caption{Training stages and loss weights.}\label{tab:stages}
        \end{table}

        To assess the contribution of codec-resynthesis supervision, we train two matched variants. \textit{DeltaMark (waveform-only augmentation; WFA)} uses only waveform attacks, whereas \textit{DeltaMark} additionally applies an EnCodec encoder--decoder cycle without RVQ during robustness training. We omit RVQ in the training attack to retain differentiability and allow gradients to propagate through the codec transformation.
    
        During robustness training, one stochastic waveform distortion is sampled per example from white or pink noise, gain, filtering, resampling, clipping, dropout, smoothing, and echo~\cite{AudioSeal,WavMark}. Attack sampling is adaptively reweighted toward lower-performing ones.

    \subsection{Evaluation setup}\label{sec:evaluation_setup2}
        Robustness is evaluated on 2,000 clips from the LibriSpeech \texttt{test-clean} split, processed using the training preprocessing pipeline and used consistently across all methods and transformations.
        
        We compare against AudioSeal v0.2.0~\cite{AudioSeal} and WavMark v0.0.3~\cite{WavMark} using their official pretrained implementations. Since our extractor is trained only for payload recovery, we fit a lightweight binary MLP detection head for 10 epochs on its frozen pooled representation. The detector threshold is calibrated on validation data to target a 1\% attack-weighted FPR and fixed across all test conditions. All methods use 16-bit payloads and the same evaluation transformations. AudioSeal and WavMark embedding and extraction are performed at 16~kHz; their watermarked audio is resampled to 24~kHz for the common evaluation transformations and back to 16~kHz before extraction.
        
        Evaluation includes the same waveform attack types used during training, evaluated under harsher conditions, repeated 24~kbps and single-pass 12--1.5~kbps EnCodec resynthesis, unseen 44.1~kHz DAC~\cite{DAC} with 7--9 quantizers, TiCodec-1g4r~\cite{TiCodec}, MP3~\cite{MP3} at 32~kbps, and AAC~\cite{AAC} at 64~kbps.
        
        We report two metrics. Detection performance is measured by the true positive rate (TPR) at the calibrated false positive rate (FPR), which quantifies whether the presence of a watermark can be established. Full-message accuracy (FMA), defined as the fraction of watermarked clips that are both detected and decoded with all 16 payload bits correct, quantifies end-to-end payload recovery and serves as the common success criterion across methods. For WavMark, detection is defined by an exact match of its fixed 16-bit synchronization pattern; for AudioSeal, the native detector output is used.
        
        Perceptual quality is evaluated using SI-SNR~\cite{SISNR}, PESQ~\cite{PESQ}, STOI~\cite{STOI}, and ViSQOL~\cite{ViSQOL} relative to the original waveform. We additionally report the unwatermarked no-RVQ EnCodec reconstruction, $x_c=D(E(x))$.

        To investigate why DeltaMark remains recoverable after EnCodec resynthesis, we analyze whether the latent perturbation itself is preserved through decoding and subsequent re-encoding.  Let $z=E(x)$ be the clean latent and let
        $d$ be the watermark perturbation. Without RVQ, we compute
        \mbox{$z'_0=E(D(z))$} and
        \mbox{$z'_d=E(D(z+d))$}. In either case, we define the recovered perturbation as
        \mbox{$d'=z'_d-z'_0$}. Direction preservation is measured by the projection ratio \(\langle d',d\rangle/\|d\|_2^2\). We compare the learned watermark perturbations with equal-norm Gaussian, temporally shuffled, and channel-shuffled controls. The analysis is performed for encoder--decoder cycling without RVQ and for EnCodec quantization/resynthesis at representative bitrates.

\section{Results and discussion}\label{sec:results}
    \subsection{Neural codec robustness}
        \subsubsection{Robustness to EnCodec resynthesis}\label{sec:codec}

            \begin{table}[h]
                \centering
                \ninept
                \setlength{\tabcolsep}{1.5pt}
                \renewcommand{\arraystretch}{0.95}
                
                \begin{tabular}{
                @{}l
                cc
                @{\hspace{1.5pt}\vrule width 0.3pt\hspace{1.5pt}}
                cc
                @{\hspace{1.5pt}\vrule width 0.3pt\hspace{1.5pt}}
                cc@{}
                }
                \toprule
                
                & \multicolumn{2}{c}{\textbf{DeltaMark}}
                & \multicolumn{2}{c}{\textbf{DeltaMark (WFA)}}
                & \multicolumn{2}{c}{\textbf{AudioSeal}} \\
                
                \cmidrule(lr){2-3}
                \cmidrule(lr){4-5}
                \cmidrule(lr){6-7}
                
                \textbf{Setting}
                & FMA & \tf{TPR}{FPR}
                & FMA & \tf{TPR}{FPR}
                & FMA & \tf{TPR}{FPR} \\
                
                \midrule
                
                24 kbps $\times$1
                & \textbf{84.0} & \tf{99.9}{2.2}
                & 63.1 & \tf{99.9}{0.4}
                & 65.0 & \tf{100.0}{0.5} \\
                
                24 kbps $\times$2
                & \textbf{60.6} & \tf{98.5}{2.3}
                & 24.4 & \tf{97.4}{0.9}
                & 20.4 & \tf{88.4}{0.1} \\
                
                24 kbps $\times$3
                & \textbf{40.4} & \tf{94.9}{2.7}
                & 11.0 & \tf{92.0}{1.5}
                & 4.0 & \tf{39.3}{0.1} \\
                
                24 kbps $\times$4
                & \textbf{24.6} & \tf{89.3}{2.8}
                & 4.4 & \tf{87.0}{2.7}
                & 0.6 & \tf{11.2}{0} \\
                
                \midrule
                
                12 kbps $\times$1
                & \textbf{75.8} & \tf{99.4}{1.2}
                & 47.0 & \tf{99.1}{0.4}
                & 19.8 & \tf{97.9}{0.5} \\
                
                6 kbps $\times$1
                & \textbf{37.0} & \tf{92.5}{1.5}
                & 13.1 & \tf{90.5}{0.3}
                & 0.1 & \tf{60.5}{3.0} \\
                
                3 kbps $\times$1
                & \textbf{4.1} & \tf{54.6}{1.3}
                & 1.1 & \tf{51.2}{0.4}
                & 0.0 & \tf{41.5}{5.3} \\
                
                1.5 kbps $\times$1
                & 0.0 & \tf{7.3}{1.4}
                & 0.1 & \tf{5.6}{0.3}
                & 0.0 & \tf{26.9}{7.6} \\
                
                \bottomrule
                \end{tabular}
                
                \caption{
                Robustness to EnCodec resynthesis (\%).
                $\times N$ denotes the number of consecutive resynthesis passes.
                WavMark produced no detections under any of the shown conditions and is omitted.
                }
                \label{tab:encodec_robustness}
            \end{table}
            
            Under increasingly severe EnCodec resynthesis, DeltaMark degrades more gradually than the waveform baselines (Table~\ref{tab:encodec_robustness}). After four 24~kbps passes, DeltaMark retains 89.3\% TPR and 24.6\% FMA, whereas AudioSeal falls to 11.2\% TPR and 0.6\% FMA, and WavMark produces no detections. Reducing the EnCodec bitrate shows the same overall pattern: DeltaMark remains robust through 12~kbps and retains 37.0\% FMA with 92.5\% TPR at 6~kbps. At 3~kbps, FMA falls to 4.1\% while TPR remains 54.6\%, suggesting that increasingly aggressive codec reconstruction first corrupts payload recovery before eliminating the watermark signal entirely.
            
            The \textit{DeltaMark (WFA)} variant follows the same general trend despite receiving no EnCodec resynthesis during training. In particular, it retains 87.0\% TPR after four 24~kbps passes and 90.5\% TPR at 6~kbps, although its FMA is substantially lower than that of the codec-trained DeltaMark model. Codec-aware training improves FMA across all non-degenerate EnCodec settings, while detection remains strong for both variants under repeated resynthesis and moderate bitrate reduction. The persistence of \textit{DeltaMark (WFA)} nevertheless suggests that EnCodec robustness is not created solely by attack-specific supervision; codec exposure instead appears to strengthen a robustness property already present in our latent formulation.

        \subsubsection{Robustness to unseen codecs}\label{sec:cross_codec}
            \begin{table}[h]
                \centering
                \ninept
                \setlength{\tabcolsep}{1.5pt}
                \renewcommand{\arraystretch}{1.00}
                
                \begin{tabular}{
                @{}l
                cc
                @{\hspace{1.5pt}\vrule width 0.3pt\hspace{1.5pt}}
                cc
                @{\hspace{1.5pt}\vrule width 0.3pt\hspace{1.5pt}}
                cc@{}
                }
                \toprule
                
                & \multicolumn{2}{c}{\textbf{DeltaMark}}
                & \multicolumn{2}{c}{\textbf{AudioSeal}}
                & \multicolumn{2}{c}{\textbf{WavMark}} \\
                
                \cmidrule(lr){2-3}
                \cmidrule(lr){4-5}
                \cmidrule(lr){6-7}
                
                \textbf{Codec}
                & FMA & \tf{TPR}{FPR}
                & FMA & \tf{TPR}{FPR}
                & FMA & \tf{TPR}{FPR} \\
                
                \midrule
                
                DAC q7
                & \textbf{56.5} & \tf{96.5}{0.3}
                & 48.3 & \tf{55.2}{0}
                & 0.0 & \tf{0}{0} \\
                
                DAC q8
                & 67.9 & \tf{98.7}{0.3}
                & \textbf{73.2} & \tf{76.7}{0}
                & 0.0 & \tf{0}{0} \\
                
                DAC q9
                & 76.1 & \tf{99.5}{0.4}
                & \textbf{89.9} & \tf{91.0}{0}
                & 0.0 & \tf{0}{0} \\
            
                TiCodec-1g4r
                & 0.0 & \tf{6.9}{0.2}
                & 0.0 & \tf{0}{0}
                & 0.0 & \tf{0}{0} \\
                
                \midrule
                
                MP3 32 kbps
                & 96.5 & \tf{100}{0.6}
                & \textbf{100} & \tf{100}{0}
                & 86.9 & \tf{96.7}{0} \\
                
                AAC 64 kbps
                & 96.6 & \tf{100}{1.0}
                & \textbf{100} & \tf{100}{0}
                & 99.8 & \tf{100}{0} \\
                
                \bottomrule
                \end{tabular}
                
                \caption{
                    Robustness to unseen codecs, in \%.
                    DAC, TiCodec, MP3, and AAC are not used during training.
                }
                \label{tab:cross_codec}
            \end{table}
            
            Under conventional lossy compression, all methods remain highly robust (Table~\ref{tab:cross_codec}). Detection is saturated for DeltaMark and AudioSeal under MP3 and AAC, and AudioSeal attains marginally higher FMA (100\% versus 96.5--96.6\%).

            Transfer to unseen neural codecs is more differentiated. Across all DAC settings, DeltaMark remains reliably detectable, with TPR between 96.5\% and 99.5\%, and retains FMA between 56.5\% and 76.1\% despite DAC not being used during training. Its robustness is therefore not restricted to the exact EnCodec transformation used during optimization. AudioSeal achieves higher FMA than DeltaMark at DAC q8 and q9, although its TPR remains lower. A possible explanation is that AudioSeal is trained with EnCodec augmentation~\cite{AudioSeal}, which may promote transfer to related neural-codec distortions.
            
            This transfer does not extend to TiCodec-1g4r, under which DeltaMark detection is reduced to 6.9\% at 0.2\% FPR and no complete payload is recovered, while AudioSeal and WavMark produce no detections. Robustness to learned codec reconstruction is thus strongly codec- and method-dependent.

    \subsection{Robustness to standard waveform distortions}\label{sec:attacks}
           \begin{table}[h]
            \centering
            \ninept
            \setlength{\tabcolsep}{1.30pt}
            \renewcommand{\arraystretch}{1.0}
            
             \resizebox{\columnwidth}{!}{%
            \begin{tabular}{
            @{}ll
            cc
            @{\hspace{1.5pt}\vrule width 0.3pt\hspace{1.5pt}}
            cc
            @{\hspace{1.5pt}\vrule width 0.3pt\hspace{1.5pt}}
            cc@{}
            }
            \toprule
            
            & & \multicolumn{2}{c}{\textbf{DeltaMark}}
            & \multicolumn{2}{c}{\textbf{AudioSeal}}
            & \multicolumn{2}{c}{\textbf{WavMark}} \\
            
            \cmidrule(lr){3-4}
            \cmidrule(lr){5-6}
            \cmidrule(lr){7-8}
            
            \textbf{Attack} & \textbf{Setting}
            & FMA & \tf{TPR}{FPR}
            & FMA & \tf{TPR}{FPR}
            & FMA & \tf{TPR}{FPR} \\
            
            \midrule
            
            --
            & --
            & 97.4 & \tf{100}{1.1}
            & \textbf{100} & \tf{100}{0}
            & \textbf{100} & \tf{100}{0} \\
            
            White noise
            & 20 dB
            & \textbf{61.5} & \tf{99.7}{15.1}
            & 34.8 & \tf{52.0}{0}
            & 1.5 & \tf{6.6}{0} \\
            
            Pink noise
            & 20 dB
            & \textbf{73.2} & \tf{99.3}{3.4}
            & 29.1 & \tf{48.3}{0}
            & 16.4 & \tf{33.9}{0} \\
            
            Gain
            & $0.1\times$
            & 58.8 & \tf{99.9}{3.3}
            & 35.3 & \tf{87.1}{0}
            & \textbf{98.9} & \tf{99.8}{0} \\
            
            Low-pass
            & 0.75 kHz
            & 36.6 & \tf{95.4}{0.2}
            & 3.0 & \tf{99.5}{0}
            & \textbf{71.8} & \tf{88.1}{0} \\
            
            High-pass
            & 1.5 kHz
            & 13.8 & \tf{100}{2.6}
            & 0.5 & \tf{76.1}{0}
            & \textbf{100} & \tf{100}{0} \\
            
            Resampling
            & 32 kHz
            & 97.4 & \tf{100}{1.1}
            & \textbf{100} & \tf{100}{0}
            & \textbf{100} & \tf{100}{0} \\
            
            Clipping
            & thr.\ 0.5
            & 97.1 & \tf{100}{1.1}
            & \textbf{100} & \tf{100}{0}
            & \textbf{100} & \tf{100}{0} \\
            
            Dropout
            & frac.\ 0.2
            & 91.6 & \tf{100}{1.2}
            & \textbf{100} & \tf{100}{0.1}
            & 97.1 & \tf{99.1}{0} \\
            
            Smoothing
            & kernel 12
            & 11.1 & \tf{99.6}{8.8}
            & \textbf{99.5} & \tf{100}{0}
            & 96.6 & \tf{98.6}{0} \\
            
            Echo
            & delay 8000
            & 83.1 & \tf{100}{3.7}
            & \textbf{99.8} & \tf{99.8}{0}
            & 64.0 & \tf{92.0}{0} \\
            
            \bottomrule
            \end{tabular}
            
            }
            
            \caption{
                Robustness to conventional waveform distortions, in \%.
            }
            \label{tab:waveform}
        \end{table}

        Table~\ref{tab:waveform} shows that DeltaMark retains broad robustness to conventional waveform distortions. Detection remains above 99\% TPR for all attacks except low-pass filtering (95.4\%). FMA is more attack-dependent, remaining high under resampling, clipping, dropout, and echo but dropping under noise, gain changes, and filtering. Overall, our formulation maintains broad robustness to conventional waveform distortions alongside its resilience to neural codec resynthesis.
    
        No method dominates across all attacks. DeltaMark retains higher FMA under additive noise, although this is accompanied by substantially higher FPR, while WavMark performs better under gain and filtering, and both baselines are markedly stronger under smoothing. The gap between detection and payload recovery is especially clear under high-pass filtering (100\% TPR vs.\ 13.8\% FMA), while white noise causes the largest increase in FPR to 15.1\%. This illustrates the value of reporting detection and exact message recovery separately.

    \subsection{Perceptual impact}\label{sec:quality}
        \begin{table}[h]
            \centering
            \normalsize
            \setlength{\tabcolsep}{2pt}
            \renewcommand{\arraystretch}{0.9}
            \begin{tabular}{lcccc}
                \toprule
                Method
                & SI-SNR $\uparrow$
                & PESQ $\uparrow$
                & STOI $\uparrow$
                & ViSQOL $\uparrow$ \\
                \midrule
        
                DeltaMark
                & 8.76
                & 3.32
                & 0.969
                & 4.29 \\
        
                EnCodec (no RVQ)
                & 9.07
                & 3.89
                & 0.978
                & 4.36 \\
        
                \midrule
        
                AudioSeal
                & 27.24
                & \textbf{4.48}
                & \textbf{0.998}
                & \textbf{4.41} \\
        
                WavMark
                & \textbf{33.71}
                & 4.26
                & 0.997
                & 4.39 \\
        
                \bottomrule
            \end{tabular}
        
            \caption{
            Perceptual quality relative to the original audio.
            EnCodec (no RVQ) denotes the corresponding unwatermarked
            encoder--decoder reconstruction.
            }
            \label{tab:quality}
        \end{table}

        Relative to the original waveform, DeltaMark exhibits lower objective quality than the waveform-domain baselines (Table~\ref{tab:quality}). This end-to-end comparison, however, confounds two sources of distortion, namely the frozen no-RVQ EnCodec reconstruction and the watermark itself. The unwatermarked reconstruction $x_c = D(E(x))$ alone yields an SI-SNR of 9.07~dB and a PESQ of 3.89, close to the values obtained for the watermarked output. To isolate the contribution of watermark insertion, we compare $x_w$ directly with $x_c$. This carrier-referenced evaluation yields an SI-SNR of 15.52~dB, a PESQ of 3.73, a STOI of 0.985, and a ViSQOL of 4.58, indicating that a substantial part of the end-to-end degradation is attributable to the codec reconstruction rather than to the watermark. Future work should therefore prioritize a higher-fidelity latent carrier while preserving the observed robustness to neural codec resynthesis.

    \subsection{Structure of latent perturbations}
        \begin{table}[h]
            \centering
            \normalsize
            \setlength{\tabcolsep}{4pt}
            \renewcommand{\arraystretch}{0.8}
            \begin{tabular}{lccc}
                \toprule
                Direction
                & No RVQ
                & EnCodec 24 kbps
                & EnCodec 6 kbps \\
                \midrule
                Learned
                & \textbf{0.528}
                & \textbf{0.458}
                & \textbf{0.375} \\
        
                Time shuffle
                & 0.496
                & 0.416
                & 0.323 \\
        
                Channel shuffle
                & 0.305
                & 0.246
                & 0.174 \\
        
                Gaussian
                & 0.308
                & 0.242
                & 0.158 \\
                \bottomrule
            \end{tabular}
            \caption{
                Preservation of learned and control latent perturbations through EnCodec.
                Entries report the mean projection ratio of the recovered perturbation onto its
                original direction. All controls are norm-matched to the learned perturbation.
            }
            \label{tab:direction_preservation}
        \end{table}
        
        A neural codec is designed to discard latent components that are unnecessary for reconstruction. This means a watermark can only survive resynthesis if it occupies directions that the codec retains. Table~\ref{tab:direction_preservation} shows that an arbitrary perturbation is largely removed, as confirmed by the Gaussian control, whereas learned perturbations are preserved substantially more strongly, with mean projection ratios of 0.458 versus 0.242 at 24~kbps and 0.375 versus 0.158 at 6~kbps. The same separation is present in the no-RVQ cycle, suggesting that it is not caused solely by quantization. The embedder has thus identified directions that the codec treats as signal rather than noise, which explains the robustness observed without codec-aware training. 
        
        The shuffled controls indicate which structure contributes to this preservation. Temporal shuffling retains most of the learned projection, with ratios of 0.416 at 24~kbps and 0.323 at 6~kbps. In contrast, channel shuffling reduces preservation to 0.246 and 0.174, close to the Gaussian control. This indicates that the advantage of the learned perturbations depends more strongly on how perturbation energy is allocated across latent channels than on its temporal arrangement.

    \section{Conclusion}
        Our latent watermark formulation degrades more gradually than waveform-domain methods under repeated and low-bitrate EnCodec resynthesis, remains detectable under the unseen DAC codec, and retains broad robustness to conventional distortions. A substantial part of this robustness is present without codec-aware training, which strengthens rather than creates it. The perturbation analysis provides a mechanism for this behavior. Learned perturbations survive codec cycling considerably better than equal-norm random controls, and disrupting their channel organization is far more damaging than disrupting their temporal structure, indicating that robustness is linked to channel-structured latent directions that the codec preferentially preserves.

        Several limitations remain. The embedding and attack codec coincide, so the EnCodec results represent a favorable case, and robustness does not generalize to all neural codecs, as the failure under TiCodec shows. Perceptual quality is bounded by the frozen codec reconstruction and remains below that of the waveform baselines. The content-independent design implies that clips carrying the same payload share the same perturbation, which is vulnerable to estimation and removal in an adversarial setting, so practical deployment would require keyed or content-dependent embedding. Future work should therefore address such adaptive latent embedding, higher-fidelity carriers, and the question of which codec representations preserve watermark-aligned directions.

    \subsection*{Compliance with Ethical Standards}
    \vspace{-0.4em}
    This study used the publicly available LibriSpeech corpus and involved no new human-subject data collection. No ethical approval was required.
    
    \vspace{-0.6em}
    \subsection*{Acknowledgments}
    \vspace{-0.4em}
    No funding was received for conducting this study. The authors have no relevant financial or nonfinancial interests to disclose.
    \vspace{-0.6em}
\bibliographystyle{IEEEbib}
\bibliography{Latent_Audio_Watermarking}
\end{document}